\documentstyle[11pt,moriond,epsf]{article}

\def\be{\begin{equation}}
\def\ee{\end{equation}}
\def\bea{\begin{eqnarray}}
\def\eea{\end{eqnarray}}

\begin{document}
%\vspace*{1cm}
\title{INTERMEDIATE REGIME BETWEEN THE FERMI GLASS AND 
THE MOTT INSULATOR IN ONE DIMENSION}

\author{DIETMAR~WEINMANN$^{1}$, JEAN-LOUIS~PICHARD$^{2}$, 
PETER~SCHMITTECKERT$^{2,3}$, RODOLFO~A.~JALABERT$^{3}$}

\address{$^1$ Institut f\"ur Physik, Universit\"at Augsburg,
           86135 Augsburg, Germany \\
	   $^2$ CEA, Service de Physique de l'Etat Condens\'e,
           Centre d'Etudes de Saclay, F-91191 Gif-sur-Yvette, France \\
           $^3$ Institut de Physique et Chimie des Materiaux de Strasbourg,
           23 rue de Loess, F-67037 Strasbourg, France}
\maketitle

\abstracts{
We consider the ground state reorganization driven by an increasing nearest 
neighbor repulsion $U$ for spinless fermions in a strongly disordered ring. 
When $U \rightarrow 0$, the electrons form a glass with Anderson localized 
states. At half filling, a regular array of charges (Mott insulator) is pinned 
by the random substrate when $U\rightarrow \infty$. Between those two 
insulating limits, we show that there is an intermediate regime where the 
electron glass becomes more liquid before crystallizing. The liquid-like 
behavior of the density-density correlation function is accompanied by an 
enhancement of the persistent current. 
}

\section{Introduction}
The interplay between the quantum interferences due to a random substrate and 
the correlations induced by charge repulsion is a central problem of mesoscopic
physics. Its understanding seems to be a necessary step towards explaining the 
recently discovered metallic phase in two  dimensions~\cite{kravchenko} and the
large value of the persistent currents observed in mesoscopic 
rings~\cite{moriond96}. In one dimension (1d), assuming a Luttinger liquid in 
the clean limit and using renormalization group arguments for including elastic
scattering, one finds~\cite{gs} that repulsive interactions cannot delocalize 
spinless fermions. However, two particles in a disordered chain can be 
(de)localized~\cite{shepelyansky} on a length $L_2$ larger than the 
one-particle localization length $L_1$ when a short range interaction is 
present. This is valid for two particle excitations of sufficient energy while 
a ground state built from two one-particle states localized far away from each 
other remains obviously unchanged. Interesting effects for the ground state 
require reasonably high filling factors. This led us to study the ground state 
of half filled strongly disordered rings and to find that the conclusions of 
Ref.~\cite{gs} are no longer valid when $L_1$ becomes of the order of the 
separation between the carriers. In this case, charge repulsion gives rise to 
successive reorganizations of the ground state, accompanied by a substantial 
delocalization of the particles and a large enhancement of the persistent 
current~\cite{schmitteckert}. The interactions $U_{\rm c}$ for which the 
charge density is reorganized fluctuate from sample to sample. Thus, the 
delocalization effect becomes essentially negligible, though visible after 
ensemble averaging. But it becomes very striking if one studies the orbital 
response of individual mesoscopic samples as a function of $U$.
 
 At a finite density and when $U \rightarrow 0$, the electrons form a glass 
with Anderson localized states filled up to the Fermi energy. This Fermi glass 
is characterized by an inhomogeneous charge configuration imposed by the random
substrate. When $U \rightarrow \infty$, at half filling for a nearest neighbor 
repulsion $U$, the electrons form a Mott insulator (or a pinned Wigner crystal 
for long range Coulomb repulsion and lower filling factors). It is 
characterized by large crystalline domains, possibly separated by some 
remaining dislocations which vanish as $U \rightarrow \infty$. The crossover 
between those two different insulators yields successive spatial 
reorganizations of the ground state~\cite{schmitteckert} which correspond to 
weakly avoided crossings between the ground state and the first excitation. 
At those level crossings, the system becomes more sensitive to any external 
perturbation as a twist in the boundary conditions or a flux through a 
ring, and this enhances the persistent currents.

\section{Model and method}

We consider $N$ spinless fermions on a chain of $L$ sites with nearest 
neighbor (NN) interaction
\begin{equation}
H= -t\sum_{i=1}^{L} (c_i^{\dagger} c_{i-1} + c^{\dagger}_{i-1}c_i) 
+\sum_{i=1}^{L} V_i n_i + U \sum_{i=1}^{L} n_i n_{i-1}
\end{equation}
and twisted boundary conditions, $c_0=\exp({\rm i}\Phi) c_L$. The operators 
$c_i$ ($c^{\dagger}_{i}$) destroy (create) a particle on site $i$ and 
$n_i=c^{\dagger}_ic_i$ is the occupation operator. The on-site random 
energies $V_i$ are drawn from a box distribution of width $W$. The strength 
of the disorder $W$ and the interaction $U$ are measured in units of the 
kinetic energy scale ($t\!=\!1$). We study interaction between NN at half 
filling, where the ground state will be a periodic array of charges located 
on the even or odd sites of the chain when $U \rightarrow \infty$. 
 The numerical results are obtained using the density matrix 
renormalization group (DMRG) algorithm~\cite{dmrg}. Ground state properties in 
disordered 1d systems can be calculated with an accuracy comparable to exact 
diagonalization, but for much larger systems~\cite{peter}.

\section{Crossover glass-liquid-crystal for increasing repulsions}

\begin{figure}[h]
\centerline{\parbox[b]{2.5in}{\epsfxsize=2.3in\epsffile{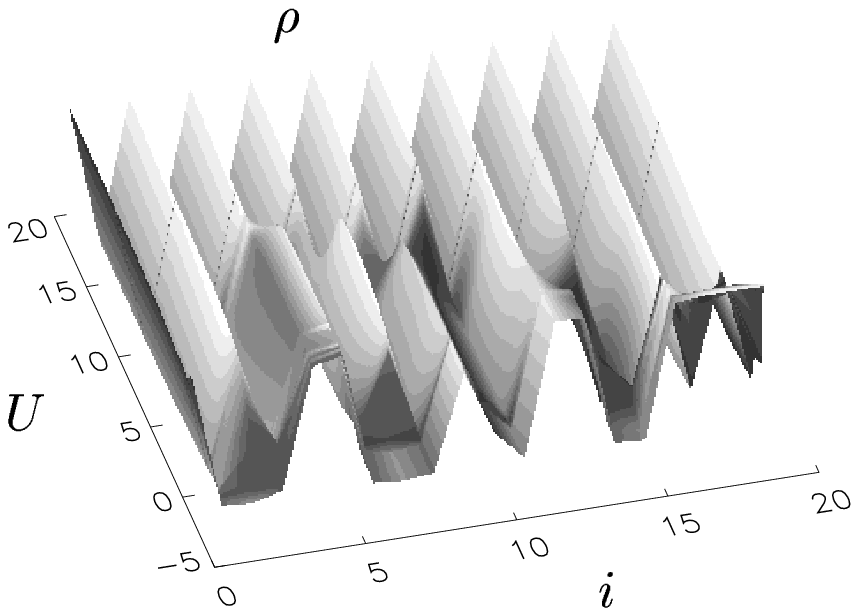}
\epsfxsize2in\epsffile{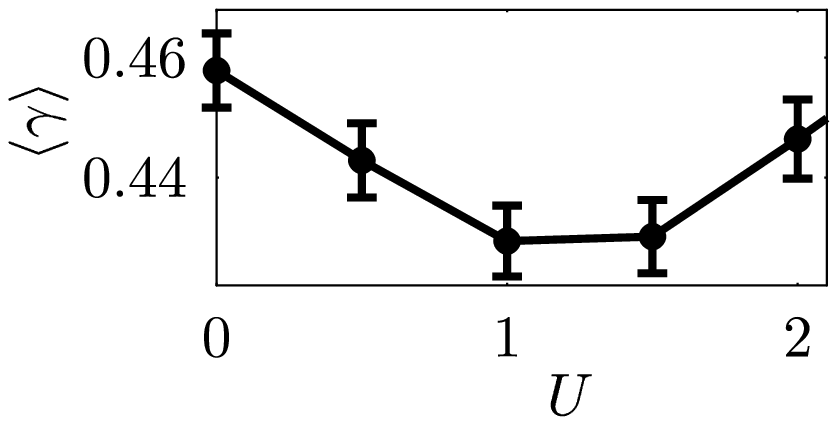}\vspace*{3mm}}
\hfill\epsfxsize=4.2in\epsffile{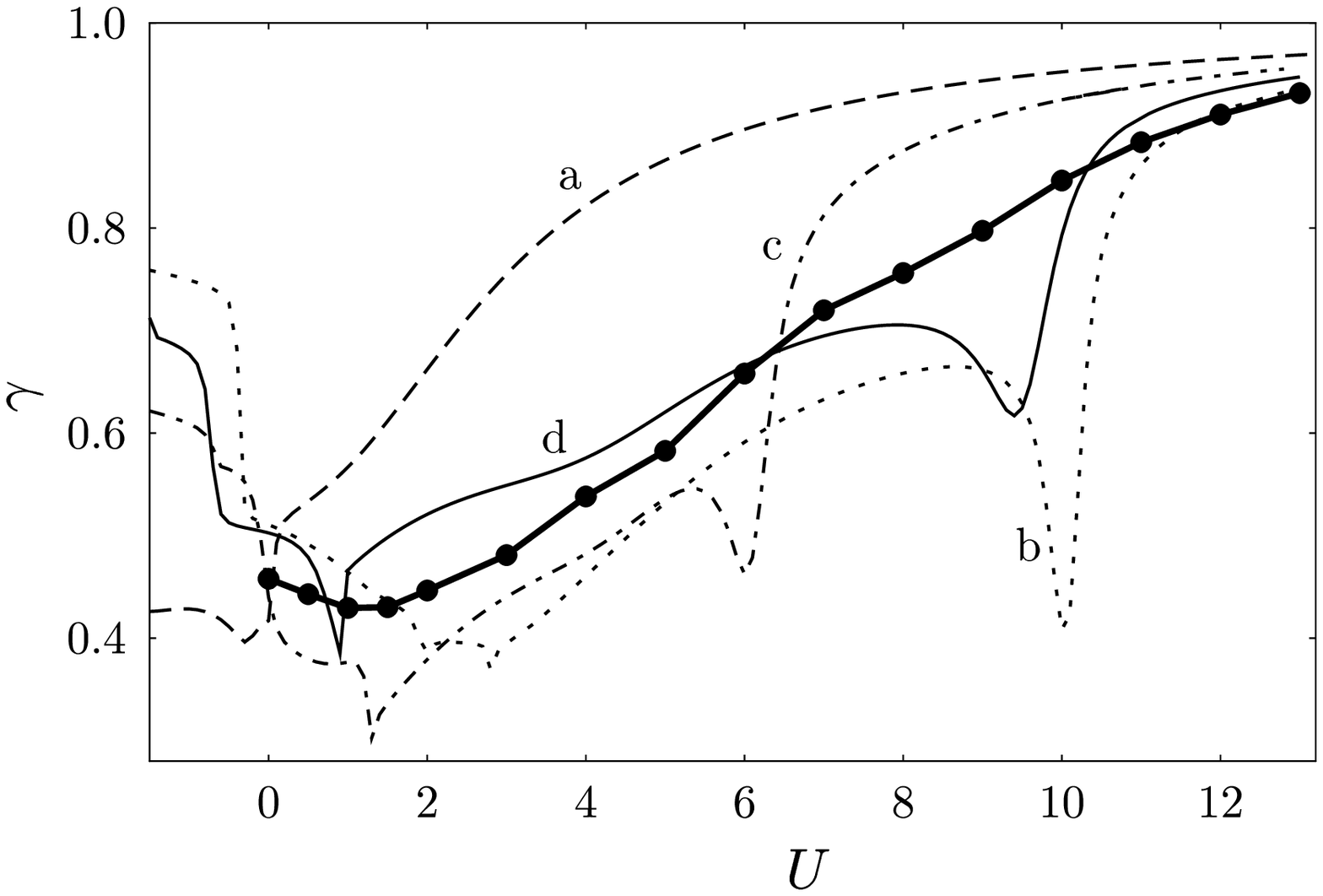}}
\caption[fig1]{\label{density} \protect\small
Top left: Charge configuration for a typical sample 
(d) for $N=10$ particles on $L=20$ sites at $W=9$.
Right: Density--density correlation parameter $\gamma$ for four samples 
with $N=10$, $L=20$ and $W=9$. Thick dots: Average over up to 166 samples.
Bottom left: Average of $\gamma$ in the region around $U=U_{\rm F}$.}
\end{figure}

 The reorganization of the ground state induced by the NN repulsion is shown 
in Fig.~\ref{density}, where the density $\rho$ (expectation value of $n_i$) 
is plotted as a function of $U$ and site index $i$. To favor the 
inhomogeneous configuration, the disorder is taken large ($W\!=\!9$) and 
$L_1\!\approx\!100/W^2$ is of order of the mean spacing 
$k_{\rm f}^{-1}\!=\!2$ between the charges. For $U\!\approx\!0$, one can see 
a strongly inhomogeneous and sample dependent density, while for large $U$ a 
periodic array of charges sets in. These two limits are separated by a sample 
dependent crossover regime.

In order to quantitatively describe this reorganization of the charge density, 
we calculate the density--density correlation function 
\begin{equation}
C(r)=\frac{1}{N}\sum_{i=1}^{L}\rho_i\rho_{i+r}
\end{equation}
for values $0\leq r \leq L/2$. The parameter $\gamma=\max_r\{C(r)\}-\min_r\{C(r)\}$
is used to distinguish between the electron liquid with constant density 
($\gamma=0$) and the regular crystalline array of charges ($\gamma=1$). If one 
includes the translation $r=0$ in the definition of $\gamma$, one gets 
$\gamma \neq 0$ for the electron glass. So $\gamma$ measures charge 
crystallization from an electron liquid as well as the melting of the 
glassy state towards a more liquid ground state.

 Fig.~\ref{density} (right) shows the dependence of $\gamma$ on the 
interaction strength $U$ for four individual samples. For certain impurity
configurations, like in sample (a), the periodic array is obtained at
a weak repulsive interaction, while one needs a strong interaction for other 
samples like (b) and (d). Typically, $\gamma$ assumes a minimum for a small 
repulsive interaction of the order of the kinetic energy scale $t$. This means
that the charge distribution is closest to a liquid there and suggests a 
maximum of the mobility of the charge carriers. This is an indication 
for a delocalization of the ground state by repulsive interactions.
In addition, most of the samples show small steps in the interval 
$0\le U \le 2t$, caused by instabilities between different configurations
of similar structure. 
 The formation of the regular array of charges imposed by strong repulsive 
interactions occurs only at an interaction strength $U\approx U_{\rm W}$ with 
$U_{\rm W}\propto W$. It can be inferred from the step-like increase of 
$\gamma$ in individual samples that the regular array is 
established abruptly at an interaction value which depends on the disorder 
realization of the given sample. Therefore, the jumps of $\gamma$ are 
smeared out when the average over the ensemble is calculated. 

Nevertheless, it can be clearly observed that the charge density does not
cross over smoothly from the disordered Fermi glass to the regular Mott 
insulator. There is a regime at weak repulsive interaction 
$U\approx U_{\rm F}\approx t$, where the interaction delocalizes the 
particles and 
counteracts the disorder before the tendency towards the Mott insulator 
starts to dominate at stronger interaction strength. 
This is similar to the behavior observed in 2d
systems~\cite{benenti_here,benenti_new}, where an intermediate phase has been 
found at moderate Coulomb repulsion and an abrupt transition to a 
regular Wigner crystal when the interaction strength increases.  

\section{Enhanced persistent currents for intermediate repulsions}

%
% phase sensitivity and avoided crossings
%

%
% ************** DEFINITION OF THE CHARGE STIFFNESS******************
%
 To measure the delocalization effect associated to the change of charge 
configuration, we study the phase sensitivity of the ground state. 
The energy difference between periodic ($\Phi\!=\!0$) and anti-periodic 
($\Phi\!=\!\pi$) boundary conditions, $\Delta E = (-)^N (E(0)\!-\!E(\pi))$ 
conveys similar information, in the localized regime, as other measures 
of the response of the ground state to an infinitesimal flux threading the 
ring: the Kohn curvature (charge stiffness) $\propto E^{''}(\Phi\!=\!0)$ 
and the persistent current~\cite{moriond96}$ J \propto -E^{'} (\Phi\!=\!0)$. 
For strictly 1d systems, the sign of $E(0)\!-\!E(\pi)$ simply depends 
on the parity of $N$, and the factor $(-)^N$ makes $\Delta E$ positive. 
\begin{figure}[tbh]
\centerline{\epsfxsize=3in\epsffile{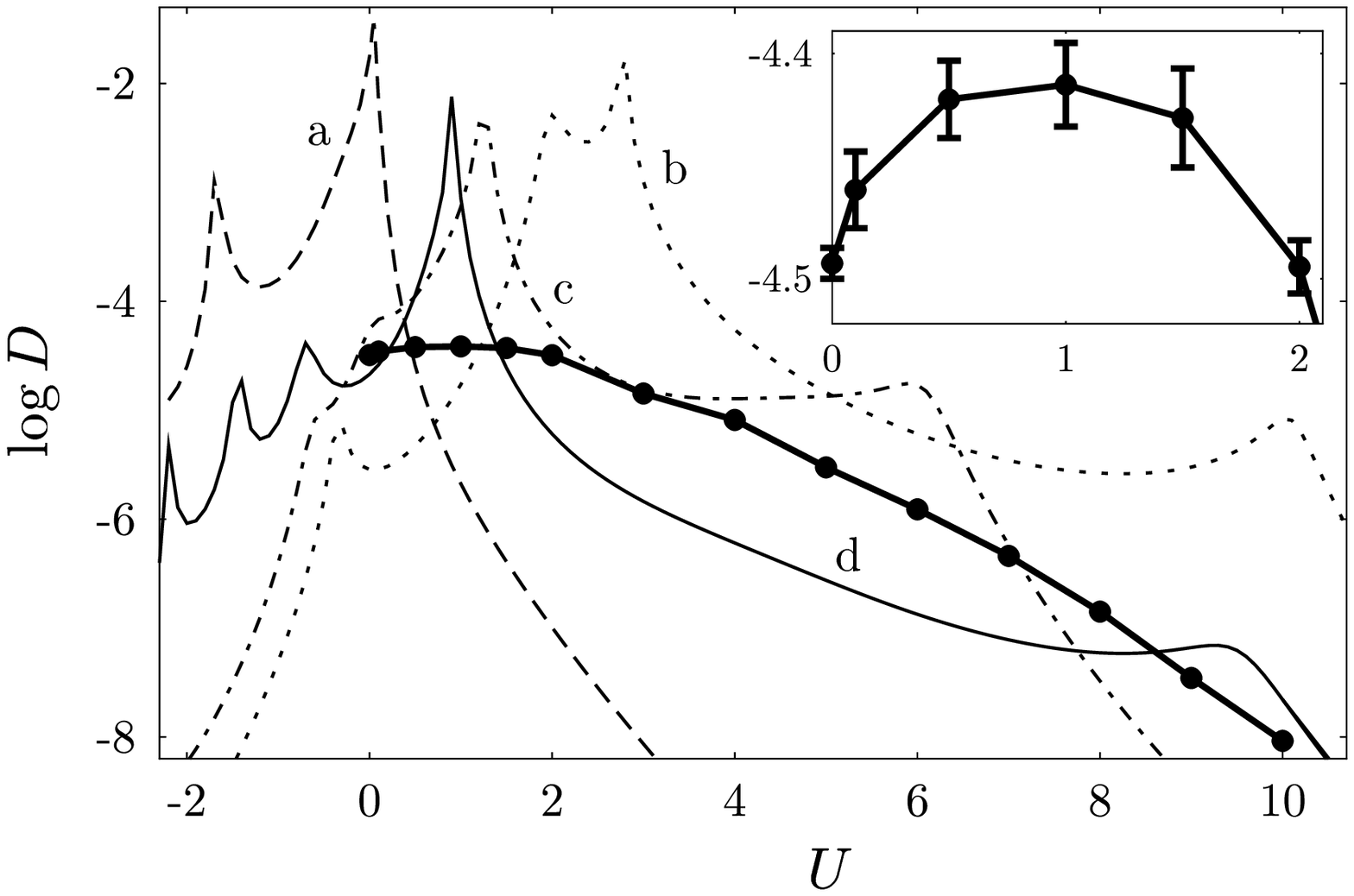}\hfill
\epsfxsize=3in\epsffile{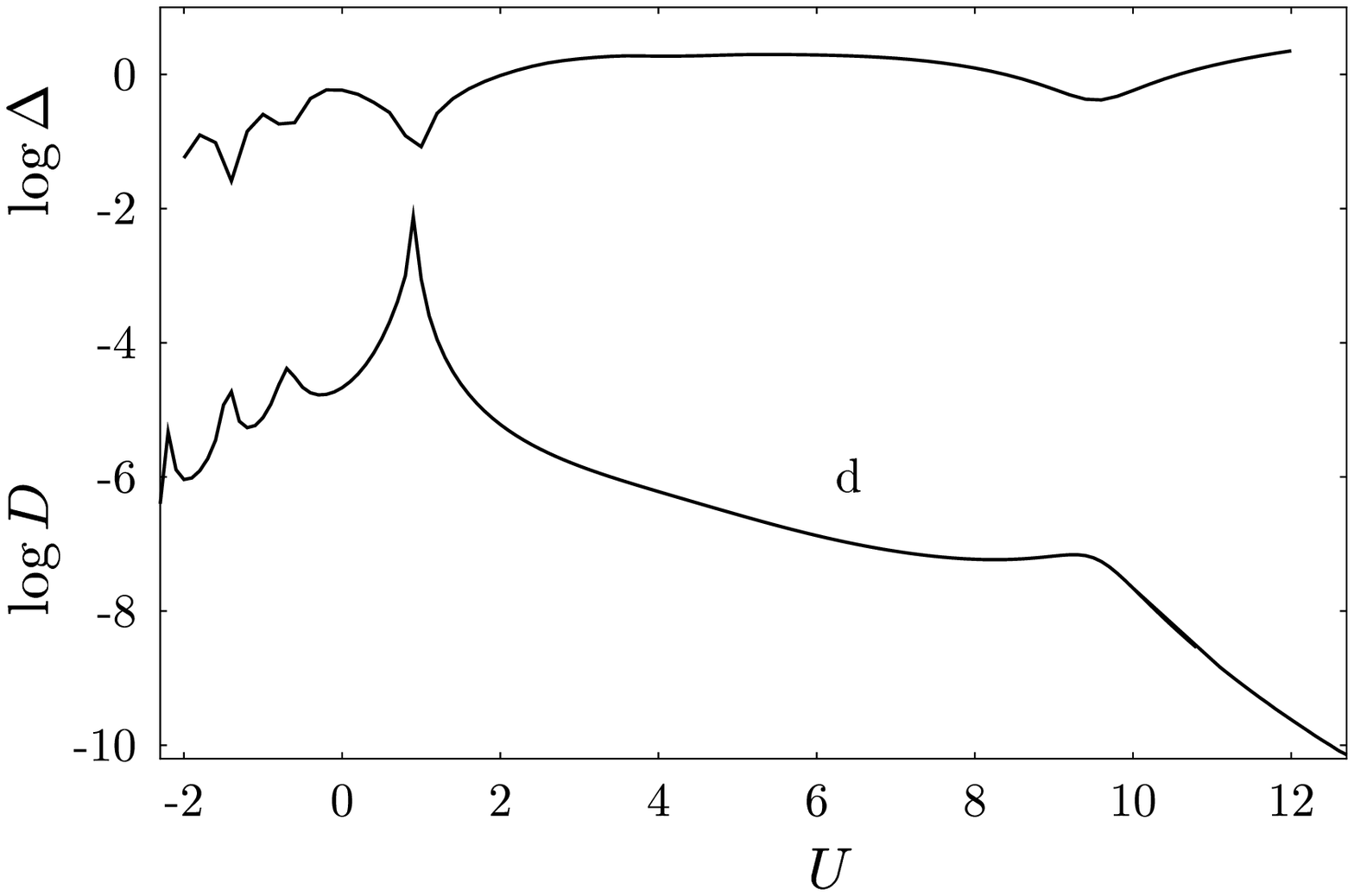}} 
\caption[fig4]{\label{m20abcd} \protect\small
Left: Phase sensitivity $D(U)$ for the four different 
samples of Fig.~\ref{density} in (decimal) logarithmic scale.
Thick dots and inset: average of $\log(D)$.
Right: The energy spacing $\Delta$ between the ground state and the first 
excited state of sample (d) (upper line) together with the corresponding 
phase sensitivity (lower line).}
\end{figure}
The phase sensitivity $D(U)=(L/2) \Delta E$ is shown in Fig.~\ref{m20abcd} 
for four samples at half filling with $W\!=\!9$. Both for 
$U\!\approx\!0$ and $U\!\gg\!1$, $D(U)$ is very small, but sharp peaks appear 
at sample dependent values $U_c$, where $D(U_c)$ in certain samples can be 4 
orders of magnitude larger than for free fermions. 
 Remarkably, the curves for each sample do not present any singularity at 
$U\!=\!0$ which could have allowed to locate the free fermion case. Peaks 
can be 
seen at different sample dependent values of $U$ (positive or negative). 
A comparison of the peak values with the step positions in Fig.~\ref{density} 
shows that the sharp peaks are in all cases accompanied by a change of the 
ground state structure.

For small repulsive 
interactions, the system is an Anderson insulator delocalized by $U$, and 
$D(U)$ increases as a function of $U$, consistent with the decrease of 
$\gamma$ seen in Fig.~\ref{density}. This decrease means that one has first a 
progressive melting of the electron glass when one turns on the interaction. 
At the maximum of D(U) occurring at the largest value of the interaction, the 
regular array of charges is established, and thereafter it becomes more and 
more rigid (pinned by the random lattice); thus $D(U)$ decreases as 
a function of $U$. 

The thresholds $U_c$ are strongly sample dependent giving rise to a very wide 
distribution of $D(U)$: the ensemble average at a given $U$ mixes very 
different behaviors and provides very incomplete information. As shown in 
Fig.~\ref{m20abcd}, $\langle\log D(U)\rangle$ decreases for repulsive 
interactions, except for a small interval around $U_{\rm F}\!\approx\!t$ 
(inset) where a local maximum is obtained. This is reflected in the behavior of
the ensemble average $\langle \gamma \rangle$ (Fig.~\ref{density}), which 
increases smoothly at large interaction except for a maximum at 
$U_{\rm F}$. 

%
% ******** LARGE DISORDER IS NEEDED AT HALF FILLING *****
%                ROLE OF THE DISORDER
%
 This delocalization effect of the ground state for $U\approx U_{\rm F}$ only 
occurs at strong disorder. For weak disorder ($W\!=\!2, L_1\!\approx\!L$) we 
recover~\cite{schmitteckert} the behavior expected starting from the clean 
limit, using bosonization and renormalization group arguments~\cite{peter,gs2}:
a repulsive interaction reinforces localization, in contrast to a (not too 
strong) attractive interaction which delocalizes. However, the conclusion that 
a repulsive interaction favors localization is no longer valid at strong 
disorder. Similar conclusions have been drawn from a study of the conductance 
of one- and two-dimensional systems at half filling~\cite{vojta}.

\section{Avoided level crossings} 
%
% avoided crossings
%
The sharp changes in the ground state structure and the peaks observed in the 
phase sensitivity are the consequences of avoided crossings between the ground 
state and the first excitation, when $U$ increases.
While the ground state at weak interaction is well adapted to the disordered
potential, another state with a different structure being better adapted to 
repulsive interactions, becomes the ground state at stronger interaction. 
In the case of large disorder, with a one-particle localization length $L_1$ 
which is of the order of the mean distance between the particles, the overlap 
matrix elements between the different noninteracting eigenstates due 
to the interaction are very small and the levels almost cross. There is only 
a very small interaction range where a significant mixing of two states is 
present. This is exactly where the peaks of the phase sensitivity appear.

This scenario is confirmed by Fig.~\ref{m20abcd} (right), where the phase 
sensitivity and the energy level spacing $\Delta$ between the ground state 
and the first excited state are shown for sample (d). 
A minimum of $\Delta$ appears always at the same interaction value 
as a peak in the phase sensitivity. A gap of increasing size opens between
the ground state and the first excitation after the last avoided crossing at 
interactions $U>U_{\rm W}$, when the Mott insulator is established. 
A study of many other samples leads to the same conclusions.
It is interesting to investigate the statistics of the first excitation energy.
In 2d with Coulomb repulsion, this has recently been addressed giving rise to
intermediate statistics~\cite{benenti_gap} at the opening of the quantum 
Coulomb gap. 

\section{Conclusions}

We summarize our main conclusions: (i) Observables for 
individual samples convey a clearer information than the (log)-averages 
over the ensemble; (ii) Spinless noninteracting fermions in 1d form an 
insulating Fermi glass at strong disorder, which becomes a Mott insulator 
when $U\approx U_{\rm W}\propto W$; (iii) At an intermediate repulsion 
$U\approx U_{\rm F}\approx t < U_{\rm W}$, the charge density is closest to a 
liquid and the mobility of the charge carriers becomes maximum, indicating 
the existence of a regime where the interaction delocalizes between 
two insulating limits. In small 2d clusters with Coulomb interaction, 
a rather similar behavior applies~\cite{benenti_here,benenti_new}. 
From a study of the longitudinal and transverse currents driven by a flux 
threading the system around its longitudinal axis, an intermediate phase 
was found, characterized by the suppression of the transverse currents 
and the persistence or the enhancement of the longitudinal currents. 
According to transport measurements, this new phase might correspond to a 
new kind of 2d metal. In one dimension, the study of $D(U)$ when the size 
$L$ increases at a given filling factor~\cite{long} allows us to exclude the 
existence of a 1d metal. We can neither characterize a change in the ground 
state by the suppression of the transverse current as in 2d. However, from a 
study of the density-density correlation functions for the ground state,  
we have shown that also in 1d, the ordered array of charges is established 
at an interaction strength $U_{\rm W}$ larger than the value $U_{\rm F}$ 
where one has a maximum interaction induced delocalization. We believe 
that the regime $ U \approx U_{\rm F}$ where the system is nor a Fermi 
glass neither a Mott insulator in 1d can be seen as a 1d precursor of 
the probable metallic phase appearing in 2d.

\section*{Acknowledgments}
We gratefully acknowledge financial support from the DAAD and the A.P.A.P.E. 
through the PROCOPE program and from the European Union through the TMR 
program.

\end{document}